# Properties of Effective Information Anonymity Regulations

*Aloni Cohen, Micah Altman, Francesca Falzon, Evangelia Anna Markatou,* and *Kobbi Nissim*[1]

[Draft: 8/23/2024]

## 1. Introduction

A firm seeks to analyze a dataset and to release the results. The dataset contains information about individual people, and the firm is subject to some regulation that forbids the release of the dataset itself. The regulation also imposes conditions on the release of the results. What properties should the regulation satisfy?

The answer depends, of course, on the scope and purpose of the regulation. We restrict our attention to regulations tailored to controlling the downstream effects of the release specifically on the individuals to whom the data relate.[2] For brevity, we call these *subject effect regulations*. A particular example of interest is an *anonymization* rule, where a data protection regulation limiting the disclosure of personally identifiable information does not restrict the distribution of data that has been sufficiently anonymized.

In this paper, we develop a set of technical requirements for anonymization rules and other subject effect regulations. The requirements are derived by situating within a simple abstract model of data processing a set of guiding general principles put forth in prior work.[3] We describe an approach to evaluating proposed subject effect regulations using these requirements --- thus enabling the application of the general principles for the design of mechanisms. As an exemplar, we evaluate competing interpretations of regulatory requirements from the EU's General Data Protection Regulation (GDPR).

Our goal is neither a comprehensive account of anonymization, nor of the principles for anonymization regulation set forth in the prior work. The technical properties we glean from the principles are few and seeming weak, leaving most of the territory uncharted. Still, we demonstrate that they are expressive enough to critically analyze prevailing interpretations of a key regulation. The result is to demonstrate how the application of the principles can lead to better clarity in examining and interpreting regulatory requirements.

---

[1] Aloni Cohen acted as first author on this paper. The other Co-Authors are listed in alphabetical order. We describe the authors' contributions following a standard taxonomy. See Liz Allen et al., Publishing: Credit Where Credit Is Due, 508 Nature 312 (2014). AC, MA, and KN contributed to the conception of the report (including core ideas and statement of research questions) and authored the first draft of the manuscript. AC led the methodology and had primary responsibility for revisions. All authors contributed to the writing through critical review and commentary. Work of K.N. was partially supported by NSF grant CCF-2217678 and by a gift to the McCourt School of Public Policy and Georgetown University.

[2] As a nonexample, a law requiring a facial recognition model to be equally accurate across many subpopulations is concerned with the downstream effects, but not specifically on the people whose data were used to train the model.

[3] Altman, Cohen, Falzon, Markatou, Nissim, Reymond, Saraogi, and Wood, *A principled approach to defining anonymization as applied to EU data protection law* (July 21, 2024). Available at SSRN: https://ssrn.com/abstract=4104748.

## 2. Six principles for anonymization

In prior work,[4] we put forward six principles for the regulation of anonymization. An anonymization rule that fails to satisfy these principles is poorly tailored to controlling disclosure of data subjects' information or the downstream effects of disclosure. These principles form the basis of the analysis in this paper. We briefly describe the six principles, and refer the reader to the prior work for elaboration.

**Process Protection** requires that data protection be evaluated based on how data is processed, rather than merely considering the result of processing. When evaluating whether a potential data release is anonymized, for example, what matters is the informational relationship to the input data. What would the release reveal about the input? This requires evaluating the input-output relationship of the data processing pipeline.

**Format Neutrality** is an elaboration of Process Protection. It requires that anonymization rules and other subject effect regulations apply regardless of data release format. The regulations should be general enough that they can be applied to a micro-data release, an interactive query system, a machine-learned model trained on the data, or any other result of data processing.

**Inclusion-Based Protection** requires defining anonymization based on how participants' information affects the release. The result of data processing should only be considered as relating to individuals who are included in the underlying dataset, and about whom the release reveals information as a result of that inclusion. This does not include information revealed about the individual that would be substantially unchanged *had the individual been excluded from the dataset* (e.g., facts about the population from which the data was drawn). Inclusion-Based Protection demands that we ask: how do the results with and without the individual differ? This principle is rooted in our focus on regulations tailored to controlling the downstream effects of the release on the data subjects specifically.

**Composition Awareness** stems from the observation that every new use or analysis of data comes with some amount of information leakage. When the same or related data is used in multiple analyses, care must be taken to control the aggregated information leakage. Composable mechanisms are those that allow one to control leakage in the aggregate by controlling leakage from individual analyses. Composition Awareness requires an anonymization regulation to permit only composable mechanisms.

---

[4] Altman, Cohen, Falzon, Markatou, Nissim, Reymond, Saraogi, and Wood, *A principled approach to defining anonymization as applied to EU data protection law* (July 21, 2024). Available at SSRN: https://ssrn.com/abstract=4104748.

**Transparency** warns against security by obscurity when evaluating the risk of a data release. Protections against disclosure or harmful downstream effects from a data release should not depend on the secrecy of the details of the processing that produced the release. If the release with the code poses an unacceptable risk to the data subjects, then the regulation should not consider the release without the code as acceptable.

**Protective Assumptions** requires that a regulation minimizes the assumptions it makes on downstream uses or users when evaluating the risk of disclosure or harm from a data release. Composition Awareness and Transparency are related to the Protective Assumptions principle. Composition Awareness stemming from the possibility that downstream users may combine two or more related data releases, and Transparency from the possibility that the details of processing come to light. The assumptions needed to discard those principles would run afoul of Protective Assumptions.

## 3. Properties of the set of mechanisms allowed by a regulation

We are interested in whether a regulation satisfies the six principles. To analyze this question rigorously requires defining 'regulation' more formally. This section sketches a formal definition of a 'regulation', within the context of data processing. We then apply this formalism to recast the principles above as properties that can be used to analyze a regulation.

We model a 'regulation' as a set of rules that impose conditions for the downstream publication and use of the results of some data analysis. For a regulation adhering to the Process Principle, the imposed conditions constrain the analysis itself. That is, the informational relationship between the inputs - which consist of information pertaining to individuals - and outputs of the analysis. The constraints may depend on the policy context (e.g., whether regulating research on scientific interventions about prisoners, reporting of routine educational assessments in public schools, or performing a machine learning analysis over patients' medical data). But within a fixed context, it is the properties of the analysis' input-output relationship that matter for controlling the downstream effects of releasing the results.

Fixing the policy context (i.e. a given target population and measures performed on it), a regulation can be viewed as specifying, implicitly or explicitly, the set of analyses whose outputs may be released. Analyses that satisfy the regulation are in the set. Analyses that don't satisfy the regulation are not.

What can we say about this set? Let us suppose that the regulation aims to control the downstream effects of publishing some results of data processing---specifically, the effects on the people to whom those data relate. If the regulation is reasonably tailored to that aim, we can already say something about the properties of the set.

## 3.1 The Process Protection principle

The first principle, *Process Protection,* states that information protection is a property of the informational relationship between what is given as input to a process and what is output from it. In other words, the process protection principle is focused on the mapping (i.e., the function) that the process implements from its input - a collection of personal information - to its output. This has a number of implications for the properties of the set.

**Non-Triviality.** First, a regulation that does not restrict the set of mechanisms has no effect. Some mechanisms should be excluded from the set, as they offer no protection. For example, a mechanism that outputs the input verbatim provides no privacy protection. We call this the *identity mechanism*. The output of the identity mechanism is the same as its input. The regulation should treat them both the same.

**At Least the Empty Release (ALTER).** Second, the set of mechanisms that the regulation considers anonymizing should not be empty. In particular, there are mechanisms whose output bears no informational relationship to its input. For example, a mechanism that ignores its input and always outputs "0", which we call the *Zero mechanism* and is equivalent to releasing no output (hence the term *empty release*), should be in the set. As the output of the Zero mechanism bears no informational relation to the input, the regulation should impose no restrictions on it.

**Not Just the Empty Release (NJER).** Third, regulations for data processing should establish some conditions under which informative data publication is allowable. That is, the set of permitted mechanisms should not consist only of the zero mechanism.

Our claims about the Identity and Zero Mechanisms are, we hope, uncontroversial. To say more about the set, we now turn to our principles.

**Assumption of Independent mechanisms**. An important assumption underlying the analysis above, and in the remainder of this section, is that the mechanisms discussed are chosen independently of any individual's specific data. Mechanisms are chosen independently of an individual's specific data if they would have remained the same under the counterfactual where the individual's information is arbitrarily changed. Mechanisms may be chosen based on properties of the population (or, more formally, the probability distribution) from which the data is drawn, such as the mean and variance, but not on the specific data.[5] For instance, if a database is deployed, then the selection of the database schema may take into consideration which attributes the data consists of without depending on their actual values in the dataset. ML-based models derived from data, like GPT-4, deviate from this assumption. In such cases, one should instead consider the machine learning process that created the model, if that process is itself independent of the data.

---

[5] Consider, for example, a process A that -- regardless of the data fed to it -- always outputs "Alice is tall". If the assumption of independence holds, then, as A's outcome has no informational relationship with its data, it provides perfect anonymization, and in particular no information about Alice is leaked by its outcome. On the other hand if the assumption of independence does not hold, then it may well be that A does leak information about Alice (or someone else). To determine whether that is the case one needs to consider the process of creating A and analyze whether that process created an informational relationship between Alice's or someone else's information and the outcome of A.

**Post processing.** The *Process Protection Principle* also gives us a way to extend a determination about one mechanism to some related mechanisms, using the concept of post-processing. Consider two mechanisms A and B. We say that "B can be computed by post-processing the outcome of A" (or "B is a post-processing of A") if the result of B can be computed given only the result of A (where 'only the result of A' means that B does not access any data other than the output of A). If B can be computed by post-processing A, then it cannot reveal more information than A does, hence, A reveals at least as much information as B.[6]

The Process Protection principle requires us to evaluate release of data outputs based *solely on what information* the output reveals about the input (rather than only the output alone as is a common alternative). This means that if the outputs of A may be released (i.e., A is in the set of analyses whose outputs may be released), then outputs of any post-processing B of A may also be released. Equivalently, if outputs of B may not be released (B is not in the set), then neither should outputs of A.

Combining post-processing with the zero mechanism implies that the output of any non-informative mechanism---one whose output bears no informational relation to its input---should be free from the regulation.[7] That is, the set contains all non-informative mechanisms. Taking this into account, NJER requires that the set of permitted mechanisms include something other than non-informative mechanisms.

This applies even to, for example, an non-informative mechanism whose output purports to be personal data (e.g., outputting "Kilroy was here." for all inputs) but in fact bears no informational relationship to anybody at all. There could be other reasons to restrict the publication of such outputs (e.g., if they are defamatory). But those reasons would be separate from the goal of controlling the downstream effects of data analysis.

Combining post-processing with the identity mechanism implies that the output of any **fully-informative mechanism**---one whose output can be used to fully reconstruct the input---should not be allowed under the regulation. That is, the set of analyses whose outputs may be released excludes all fully-informative mechanisms.

---

[6] We limit discussion here to cases where there is a computationally tractable mapping from A to B.
[7] This by design excludes indirect effects of regulation that occur solely through psychological or symbolic channels .E.g. (respectively) (a) Alice suffers emotional distress as a result of learning that her information was included in processing -- although she was unaffected by data collection, and no one learns about Alice because she was included in processing. (b) The requirement that all computing systems be made in America is interpreted as a symbol of the commitment the administration has to security and to the country.

Our remaining five principles have additional implications for what mechanisms are in the set or not.

## 3.2 The Format Neutrality principle

The *Format Neutrality Principle* elaborates the Process Principle (specifically post-processing combined with non-triviality or ALTER). It says that whether a given mechanism M is in the set of allowed mechanisms should not depend on the form of its output but on the information content of its output. Whether a mechanism is in the set should not depend on whether its output is in the form of a single number, a table, a graph, text, a picture, etc., but on the information that can be inferred from its output.

One implication is that for any space of outputs Y, there always exists mechanisms in the set of allowed mechanisms which produce outputs in Y. Namely, some post-processing of the zero mechanism.

On the other hand, for any sufficiently rich space of outputs Y, the set of allowable mechanisms excludes some mechanisms which produce outputs in Y. This is because for any Y large enough to encode personal data, there is a full-informative mechanism whose outputs are in Y.

## 3.3 The Inclusion Protection principle

The *Inclusion Protection Principle* scopes the regulation as controlling the effects of releasing a data analysis specifically on the people to whom the original data relate. Of all the principles, this is the one that most specifically relates to protecting data which relates to specific individuals from being disclosed, rather than controlling the effects of outputs of computations more generally.

As with the process principle, we envision the inclusion principle to consist of a collection of formal rules that guide how regulations should legally restrict anonymization processes. Central to our treatment is a notion of *relatedness*, which is motivated, e.g., by Article 4 of the GDPR which states that " 'personal data' means any information relating to an identified or identifiable natural person ('data subject')".

To begin, we stipulate that a data release may be related to an individual only if the individual's information was used in its making.

- **Excluded Individuals:** If an individual's data is not part of the inputs to a process, then the output of the process is not related to that individual.

We extend the above, guided by the Process Protection principle. What matters is the informational relationship between input and output. If an individual's data has no influence on the outcome of the process---whether or not their data were included in the input---then the outcome of the process should not be considered related to the individual. Data has no influence if the outcome of the process would have been the same whether or not the data was excluded or arbitrarily changed. The following is a relaxation of this understanding allowing the outcome of the process to depend very slightly on the individual's data:

- **Individual unrelatedness:** If the output of a mechanism does not depend or depends slightly[8] on a specific individual data then the output of the mechanism is unrelated to that specific individual.

Since regulations also generally speak to when data is related, we define a complementary concept:

- **Individual relatedness:** If the output of the process depends significantly on a specific individual's data --- then the output of the process is related to that individual.

Note that although the terms "depends significantly" and "depends slightly" are meant to correspond to disjoint levels of dependency, "depends slightly" is not the complement of "depends significantly". Hence, there may be a gap, or a "gray area", between individual relatedness and individual unrelatedness.

It is essential that data that is not related to an individual would remain so even if further processed. Otherwise data that is deemed anonymized and hence unregulated could turn to be regulated after being shared in public and hence effectively stripped of protection:

- **(Un)relatedness under post-processing:** If a release O is unrelated to a specific individual, then for every mapping P it is the case that P(O) is unrelated to that individual. If there exists a mapping P such that P(O) is related to a specific individual, then O is related to that individual.

### 3.3.1 Relatedness to anonymization
In applying the Inclusion Protection principle to an anonymization regulation, we must establish the connection between anonymization and relatedness to individuals. When a regulation treats a specific protection mechanism as sufficient for anonymization it necessarily implies that the regulation considers outputs under that mechanism as unrelated to any individual. In particular, by Individual Unrelatedness, if a mechanism M has the property that adding data relating to an individual to its input cannot more than slightly change its output, then M is in the set of allowed mechanisms.

However, when an output O of processing input data D is not anonymous under a regulation, and no sufficient protection criterion inherently defines O's relation to individual *i*, further details of the law must be examined in order to determine whether relatedness holds.. In general, regulations can take the approach that non-anonymized outputs are related to either: (a) all individuals whose measurements were included in the data D; (b) some individuals in the population, as a function of the output O; or (c) some individuals *i* in D, as a function of D and O.

---

[8] M is randomized computation and its output is sampled from some distribution that is a function of its input. M depends slightly on an individual's data if adding or removing the data from the input cannot more than slightly change its output distribution. M instead depends significantly if adding or removing the data from the input can significantly change its output distribution.

Only the last approach can satisfy inclusion. Consider a mechanism M that outputs the unmodified records of all males, say, in D, and nothing else. In other words, for some portion of the population (males), M acts as the Identity Mechanism. For others it acts as the Empty Release Mechanism. Outputs from M must be related to some *i* in D and unrelated to others. But (a) treats this output as related to all *i* -- even for non-males whose records cannot affect the output, violating relatedness. Option (b), on the other hand, will treat as related some *i* who are not in D and thus cannot affect O -- violating unrelatedness.

### 3.4 The Composition Awareness principle

The *Composition Awareness Principle* is a tool for reasoning about combinations of mechanisms and managing their joint effect on individual privacy. A simple version of composition is **parallel composition**. The parallel composition of mechanisms A and B is the mechanism that separately evaluates both A and B on the same data and produces both outputs.[9] In spirit, composition awareness requires that composition of allowed mechanisms (those in the set) is also allowed.[10]

### 3.5 The Transparency principle

The transparency principle is an adaptation of Kerckhoff's principle which is widely embraced in cryptography, also commonly phrased "the enemy knows the system".

**Transparent Complement.** For any mechanism A, we define a closely-related mechanism A* as follows. A* evaluates A and outputs A's result along with a detailed description of A itself (e.g., the code of mechanism A).

The Transparency Principle requires that A is in the set of mechanisms considered anonymizing only if A* is in the set. Put differently, if disclosing a detailed description of the inner workings of the mechanism A would pose a risk of harm from downstream uses of the results, then the regulation should not deem A to be in the set of analyses whose outputs may be released.

---

[9] Other forms of composition exist. Two important settings are of adaptive and concurrent composition. In an **adaptive composition** setting, mechanism A is first executed to produce some release, and the choice of mechanism B may depend on the release produced by A. **Concurrent composition** applies to query-answering mechanisms A, B which may be interrogated concurrently, so that answers given by mechanism A may influence the choice of queries made to mechanism B and vice versa.

[10] Interpreted strictly as stated, this would mean that the set of mechanisms only contains non-informative mechanisms. A more nuanced version might instead require: if the parallel composition of A and B would be patently at odds with the regulation (a la fully-informative mechanisms), then one of A or B should not be in the set. A full treatment would require introducing a numerical measure of risk, rather than just considering an allowed-disallowed binary. Composition requires that the risk from the composition of mechanisms A and B is non-trivially bounded by some function of the risks from the A and B individually. See Fluitt, Aaron, et al. "Data protection's composition problem." *Eur. Data Prot. L. Rev.* 5 (2019): 285.

### 3.6 The Protective Assumption principle

The *Protective Assumption Principle* guides us when the effects of releasing outputs of a mechanism M depend on assumptions about the downstream user or use.

If we are unwilling to make any assumptions at all, then the set will only contain non-informative mechanisms -- violating **NJER**. But we should not accept all assumptions as justification. Even releasing the output of the Identity Mechanism (i.e., the data itself) won't have any downstream effect if you assume that nobody bothers to look at the release.

**Non-Contrived Assumptions.** The *Protective Assumption Principle* (Section 2.5) only allows minimal assumptions on the downstream use as justification for a mechanism's membership in the set of analyses whose outputs may be released. A weaker condition is that no logically incompatible, facially invalid, knife-edge, or artificially-tailored assumptions are required for the regulation to satisfy the other properties defined in 3.1-3.5. This rules out, for example, assumptions that all potential adversaries lack (more than minimal) background knowledge about any individuals; or that outputs will be discoverable only within the geographic region they were published; or that measurements of individual attributes will never have extreme outliers or will always (or never) follow a normal distribution.

## 4 Applying the properties to GDPR anonymization

Regulations such as the GDPR include a variety of requirements: administrative requirements of organizations, mechanisms for enforcement, exceptions for various purposes, remediation, and more. The focus of the analysis herein is on the part of the regulation that legally controls the governing processes, methods, and measures to be used for transforming personal data into data which is not personal. Under the GDPR, such data is not regulated. That is, the focus of our analysis is on the part of the regulation that describes what is required to anonymize data under the law. With this focus, the goal of the regulation is to constrain the consequences of a data release (or multiple data releases) for the privacy of the individuals or groups whose data is processed.

In interpreting how the GDPR constrains these consequences legally the analysis derives from the text of the GDPR (in particular, Articles 1 and 4, and Recital 26) and the *Article 29 Working Party Opinion 05/2014* , which defines anonymization as protection from three general types of privacy attacks: linkability, singling out, and inference.[11]

In the remainder of this paper, we apply the properties laid out in Section 2 to these three types of privacy attacks. The findings are summarized in Table 1. In prior work[12], we analyzed the GDPR concept of singling out, demonstrated specific failures of its interpretation, and explored an alternative concept related to singling out which was designed to avoid these failures. Here, we take a somewhat different approach by systematically applying the properties of a privacy concept as constructed through these GDPR documents.

---

[11] Opinion 05/2014 (n 2), 11-12.
[12] Aloni Cohen and Kobbi Nissim, 'Towards formalizing the GDPR's notion of singling out' (2020) 117(15) Proc. Natl. Acad. Sci., 8344-8352; Micah Altman, Aloni Cohen, Kobbi Nissim and Alexandra Wood, 'What a hybrid legal-technical analysis teaches us about privacy regulation: The case of singling out' (2021) 27(1) B.U. J. Sci. & Tech. L. 1.

The Article 29 Working Party approach to defining anonymity has some very promising aspects. By breaking anonymization into protection from linkability, singling out, and inference, the working party extended the interpretation of anonymity significantly beyond its narrower traditional interpretation as protection from the (re-)identification of an individual record in a dataset where directly or indirectly identifying information was scrapped, typically by means of linkage with an identifiable dataset. Furthermore, the three concepts introduced by the working party – linkability, singling out, and inference – set a foundation for the discussion of whether the use of specific privacy enhancing technologies in sociotechnical systems handling personal information satisfy legal privacy desiderata. Lastly, the Working Party approach opened the door for a technical-mathematical modeling of the three components (some of which is described below). This modeling allows the examination of the concept of anonymity in light of the progress made in the last two decades in modeling and analyzing privacy from a technical point of view.

**Table 1: Consistency of GDPR Privacy Concepts with Minimum Formal Properties**

|  |  | Inference |  |  | singling out |  |  | linkability |  |
|---|---|---|---|---|---|---|---|---|---|
|  |  | *Dalenius* | *D&L* | *DP* | *isolation* | *k-anon* | *PSO* | *double match* | *double jeopardy* |
| **Processing** | **Non-Triviality** | yes | yes | yes | yes | yes | yes | yes | yes |
|  | **ALTER** | yes | yes | yes | no | yes | yes | yes | yes |
|  | **NJER** | no (under post processing) | yes | yes | no | yes | yes | yes | yes |
|  | **Post Processing** | yes | yes | yes | yes (vacuously)* | no | yes | no | yes |
| **Format Neutrality** | **(Implied) Format Neutrality** | yes | yes | yes | yes (vacuously)* | no | yes | no | yes |
| **Assumption Minimization** | **Non-Contrived Assumptions** | yes (vacuously)* | no (if NJER is satisfied) | yes | yes (vacuously)* | yes | yes | yes (after modification)** | yes |
| **Transparency** | **Transparent Complement** | yes (vacuously)* | yes (after modification)** | yes | yes (vacuously)* | yes | yes | yes (after modification)** | yes |
| **Composition** | **Parallel Composition** | yes (vacuously)* | unknown | yes | yes (vacuously)* | no | no | no | yes |
| **Inclusion** | **Excluded Individual** | yes (vacuously)* | no | yes | yes (vacuously)* | yes | yes | yes | yes |
| | Individual Relatedness | yes (vacuously)* | yes | yes | yes (vacuously)* | no | no | no | yes |
| | Individual Unrelatedness | yes (vacuously)* | no | yes | yes (vacuously)* | no | yes | yes | yes |

*Each table cell indicates whether required properties are satisfied by various protection concepts implied by regulation. Specifically, consistency of each property with: Delanius's definition of private inference (Delanius), Duncan & Lambert definition of private inference (D&L), Differential Privacy (DP), isolation, k-anonymity (k-anon), predicate singling out (PSO), and alternate definitions of double match linkability, and double jeopardy linkability. For details of the analysis supporting each cell, definition of properties,and definitions of protective mechanisms see Section 5-7.*

*\* "Vacuously" denoted that the property is satisfied only because the set of allowed mechanisms is empty.*
*\*\* "after modification" denotes that the regulation as written does not require the stated protection, but could be modified to explicitly do so without running afoul of other properties.*

## 5 The Concept of Inference: A Principled Analysis

Let's consider a regulation that requires anonymization to satisfy one of various competing interpretations of opinion 05/2014 definition of inference. We now examine the properties of this regulation to evaluate its consistency with the five principles.

### 5.1 The meaning of inference in Opinion 05/2014 on Anonymisation Principles

Opinion 05/2014 defines inference broadly as "the possibility to deduce, with significant probability, the value of an attribute from the values of a set of other attributes".[13] The opinion does not make its meaning precise, and expresses substantial uncertainty about whether any of the protection mechanisms examined can be used to satisfy requirements for inference protection (summarized in the Opinion's Table 6). Notwithstanding, the language, reference, and history suggest that the concept refers to some variant of "*inferential disclosure"* or (*probabilistic*) *"attribute disclosure"* within the classical disclosure control literature.

The most commonly used and referenced definition for these terms in the classical statistical disclosure limitation literature, is by Duncan and Lambert.[14] According to Duncan and Lambert, inferential disclosure occurs when an intruder reasons about the release made available by a statistical agency, together with other information, to learn (maybe without certainty) the value of a respondent-reported attribute which the statistical agency attempted to remove from the release.[15] Although the Working Party does not provide a technical definition or citation, its conceptions of inference follows similar lines.[16]

---
[13] Opinion 05/2014 (n 2), 11-12.
[14] Federal Committee on Statistical Methodology, 'Statistical Policy Working Paper 2' (1978) uses the terms 'disclosure' and 'D-disclosure' to refer to this concept, employing language that parallels Opinion 05/2014: 'If the release of the statistics S makes it possible to determine the value Dx more accurately than i., possible without access to S, a disclosure has taken place. More exactly, a D-disclosure has taken place'. Ibid. 10. They attribute this concept to Tore Dalenius, 'Towards a methodology for statistical disclosure control' (1977) 15 Statistik Tidskrift 15, 222–429. The concept was later refined, most notably by Duncan and Lambert (1989) providing a notational framework, and by Chris J. Skinner, 'On identification disclosure and prediction disclosure for microdata' (1992) 46(1) Statistica Neerlandica 21-32, who adopts an absolute threshold for prediction accuracy and refers to the modified concept as 'predictive disclosure'.

A similar desiderata was articulated by Dalenius in 1977 (also extensively cited in the statistical disclosure limitation literature): "Anything that can be learned about a respondent from the statistical database should be learnable without access to the database".[17]

The Duncan & Lambert derived definitions of inference relax the requirements of Dalenius -- they allow an adversary to learn a "small" amount about an individual as the result of the output. The amount learned is defined as a relative or absolute limit on change from prior to posterior, and/or an absolute threshold limit on the posterior. For example, under the most common variant: if Alice's prior on Bob's cancer risk was only 15% before the study, then after the study it shouldn't change by more than 10% of its original value, i.e., remain in the range 15 +/- 1.5%.

## 5.2 Applying the process principle to inference

Recall that the Process Principle requires evaluating data protection based on how data is processed. Are the competing interpretations of inference, above, consistent with this principle?

Under the stronger (Dalenius-derived) interpretation of "inference", protection must prevent any adversary, who may have no or partial information about some respondent-reported attribute, from learning more about that attribute by receiving an output of the process. As it turns out, this interpretation limits dramatically the processes that may be considered anonymizing under that definition.

Returning to an example mentioned above, a scientific study revealing a correlation between smoking and lung cancer enables this type of inference. For example, if Alice knows that Bob smokes, then from this study Alice learns something new about Bob: that he is at high risk for lung cancer (without learning whether he in fact has cancer with certainty). Hence, the scientific study would not be treated as anonymizing by Opinion 05/2014 under the Dalenius interpretation of inference.

This example extends directly to include all statistical and machine learning models enabling "deduc[ing] . . . the value of an attribute from the values of a set of other attributes".[18] Dwork and Naor generalize the example beyond smoking and cancer to essentially any useful data release.[19] They show that if Alice learns anything new from the data release, then there is some partial knowledge that she might have that reveals nothing about Bob on its own, but discloses a sensitive attribute when paired with the release.

---

[15] George Duncan and Diane Lambert, 'The risk of disclosure for microdata' (1989) 7(2) J. of Bus. & Econ. Stat. 207-217.
[16] The opinion repeatedly implies that inference is about a form of probabilistic and partial information gain, e.g. : to "limit inference" it necessary to "keep track of the queries issued by an entity and to observe the information gained about data subjects" (p 15); the main fault of k-anonymity is that it fails to control inferece (p. 17) the advantage of l-diversity over k-anonymity is that by reducing information leakage (pg 18) it reduces the confidence of inference (p 18, n2); and the more detailed examples in Appendix A2 of inferences is conistent with this conceptualization of inference.
[17] Dalenius, Tore. "Towards a methodology for statistical disclosure control." statistik Tidskrift 15, no. 429-444, 1977
[18] Opinion 05/2014 (n 2), 11-12.

In other words, the strong definition of inference allows no useful information to be released. More specifically, the Dalenius definition of inference satisfies **Non-Triviality** and **ALTER**, but violates **NJER** (under **post-processing**).

Moreover, none of the relaxations considered by Duncan and Lambert satisfy **NJER**. If the results of a data analysis cause only a small prior-to-posterior change in beliefs, as required by these relaxations, then the prior beliefs must approximate the posterior beliefs. This restriction roughly means that findings from data analyses may be disseminated only if they reinforce prior beliefs or do not change one's beliefs at all. Hence, these definitions of inference either prevent meaningful data analyses (violating **NJER**) or assume the results of those analyses are consistent with our priors (violating **non-contrived assumptions**).

We certainly do not believe that the drafters of Opinion 05/2014 intended to prohibit useful data releases altogether. For GDPR inference requirement to be meaningful it should adopt a more nuanced definition that does not consider processes as violating the requirement to protect against inference in cases where they produce "population-level" knowledge such as the correlation between smoking and cancer.

### 5.3 Applying the inclusion principle

The opinion 05/2014 definition of inference renders any attribute that may be inferred about an individual as evidence for a data release to be related to that individual. But this already is at odds with the **Excluded Individuals** principle. For example, a study revealing a connection between smoking and cancer enables inferences about smokers regardless of their inclusion in the study.

Furthermore, the opinion 05/2014 definition does not distinguish between inferences of information about a specific individual as a result of the inclusion of the individual's information in the data processing, i.e., information that could not be learned about the individual had their data not been used and other types of inferences, including inferences which apply to the entire population, such as that people who smoke are more likely to suffer lung cancer.[20] By relatedness and relatedness under post-processing, only the former type of inferences should be considered as relevant to anonymization.

---

[19] In a little more detail, Dwork and Naor show that if Alice attempting a privacy attack has some (mild) uncertainty in what the outcome of the analysis would be, then that uncertainty can be used in creating auxiliary knowledge z such that (i) by itself z does not enable Alice perform the attack, and (ii) given z and the outcome of the analysis Alice can perform the attack. See Dwork, Cynthia, and Moni Naor. 2010. "On the Difficulties of Disclosure Prevention in Statistical Databases or The Case for Differential Privacy". Journal of Privacy and Confidentiality 2 (1). https://doi.org/10.29012/jpc.v2i1.585.

[20] Taken to the extreme, the Working Party's definition of inference could be read to even include self-evident inferences such as 'people who are over 50 years old are also over 40 years old' within the definition of personal data. As argued in Section 2.5, updated guidance should clearly exclude this type of overbroad interpretation of inference.

Both the interpretation of inference by Dalenius and the weaker definition of inference by Duncan and Lambert do not satisfy the **excluded individuals** principle as a release may incur a significant change in the posterior belief about an individual, compared with the prior belief about them, whether the individual's data was or was not part of the inputs to the process generating the release.

Dalenius' interpretation of inference satisfies **individual relatedness** as when an individual's input affects the output of a mechanism significantly it is possible to learn about the individual something that would not be learnable without access to the output of the mechanism, which the Dalenius interpretation recognizes as inference. Dalenius' interpretation does not satisfy **individual unrelatedness** as the above applies even when an individual's data influences the output of the mechanism very slightly.

Similarly, Duncan and Lambert's interpretation of inference satisfies **individual relatedness**, albeit in a more meaningful sense then above: if the output of a mechanism depends significantly on an individual's data, then that would result in a significant change between some observer's prior beliefs about the individual and their posterior belief about the individual. On the other hand, a significant prior to posterior change may occur even if the dependency on the individual's data is slight or insignificant. Returning to our smoking vs. cancer example, a release may teach that an individual is more prone to lung cancer even if the individual's data affected the release very slightly. Hence, the Duncan and Lambert interpretation of inference does not satisfy **individual unrelatedness.**

## 5.4 Application of Format Neutrality, Composition, Protective Assumptions, and Transparency to Inference

On its face, inference neither requires nor prohibits format neutrality, composition awareness, protective assumption, and transparency. Surprisingly, however, common definitions of inference logically violate many of these principles. The remainder of the section shows how each of these violations occur.

Happily, any regulation that satisfies Process Protection Principle (specifically, post-processing and non-triviality) will also be format-neutral. All interpretations of inference that we consider pass this test. Explicit language requiring format neutrality is a useful reminder, but is not technically necessary.

Transparency is neither implied nor contradicted by the Duncan and Lambert definition. For example, as the opinion 05/201 notes, permutation-based (aka 'swapping'), noise-addition (referring to the classic statistical disclosure limitation techniques), and l-diversity based approaches are permissible as protections against inference -- yet these can substantially differ with respect to transparency: Most of the specific mechanisms used in these areas were developed under the implicit assumption that the mechanism would not be disclosed[21], and prominent mechanisms using these three approaches clearly fail to be protective when transparent.[22] Transparency is satisfied (vacuously) in the Dalenius interpretation of inference. Under Dalenius, the allowed mechanisms are exactly the uninformative ones (violating NJER). The transparent complement of any uninformative mechanism is still uninformative.

The Dalenius definition of inference trivially satisfies the protective assumption principle, as uninformative mechanisms are protective without needing to make assumptions (contrived or otherwise). Furthermore, as noted in section 5.2, the Duncan and Lambert (relative-risk) definition of inference is only compatible with NJER, and thus the process principle, under very strong assumptions on adversary knowledge.

One set of assumptions is explicit in the GDPR's treatment of anonymization. By Recital 26, it only requires protection against "means reasonably likely to be used" to identify a data subject. The output of a mechanism may be considered anonymized even if it is in principle identifiable, if it is only by means not reasonably likely to be used. The GDPR does not further explain what means are considered to be likely to be used and whether that is a contrived assumption or not depends how the phrase is interpreted. For example, the French data protection authority requires anonymization to make identification "impossible, in practice, ... by any means whatsoever."[23] In contrast, the UK's draft guidance does not require such absolute protection, allowing one to factor in, for example, "the likelihood of someone wanting to attempt to identify individuals", or contractual obligations.[24]

Finally, a regulation that protects against individual inferences, does not necessarily guarantee composition awareness. Generally, composition is rarely achieved without explicit design, and it is possible for a series of individually uninformative releases to be combined to provide surprising information[25] or even complete disclosure.[26] In fact, most mechanisms satisfying l-diversity, which is considered as protection for inference under opinion 05/2014, do not compose.[27] Moreover, no mechanism that composes under the Duncan and Lambert or Dalenius definitions of inference is known.[28]

---

[21] Xiao, Xiaokui, Yufei Tao, and Nick Koudas. "Transparent anonymization: Thwarting adversaries who know the algorithm." ACM Transactions on Database Systems (TODS) 35.2 (2010): 1-48.; Risk-Utility Paradigms for Statistical Disclosure Limitation: How to Think, But Not How to Act

[22] Xiao, Xiaokui, Yufei Tao, and Nick Koudas. "Transparent anonymization: Thwarting adversaries who know the algorithm." ACM Transactions on Database Systems (TODS) 35.2 (2010): 1-48.; Abowd, John M., et al. "The 2020 census disclosure avoidance system TopDown algorithm." Harvard Data Science Review Special Issue 2 (2022).; Willenborg, Leon, and Ton De Waal. Statistical disclosure control in practice. Vol. 111. Springer Science & Business Media, 1996. - Section 2.6.

[23] "L'anonymisation est un traitement qui consiste à utiliser un ensemble de techniques de manière à rendre impossible, en pratique, toute identification de la personne par quelque moyen que ce soit et de manière irréversible." https://www.cnil.fr/fr/lanonymisation-de-donnees-personnelles

[24] UK ICO, "Draft anonymisation, pseudonymisation and privacy enhancing technologies guidance" Chapter 2 (How do we ensure anonymisation is effective?) https://ico.org.uk/media/about-the-ico/documents/4018606/chapter-2-anonymisation-draft.pdf

[25] Aaron Fluitt, et al., 'Data Protection's Composition Problem' (2019) 5 Eur. Data Prot. L. Rev. 285.

[26] In fact, this property is the basis for cryptographic secret sharing schemes: Karnin, Ehud, Jonathan Greene, and Martin Hellman. "On secret sharing systems." IEEE Transactions on Information Theory 29.1 (1983): 35-41.

## 5.5 Discussion

Under close analysis, the 05/2014 definition of inference has unexpected implications -- as the analysis in sections 5.3 and 5.4 demonstrate. First, a strict enforcement of inference protection, suggested by the definition, requires that nothing can be inferred about an individual -- which is possible only when no information is released at all. Second, this definition cannot be salvaged by relaxing enforcement to allow some small trivial amount to be inferred about an individual: Even under a relaxed definition, such as that proposed by Duncan & Lambert, no output could be released, unless the adversary is assumed to have improbable and inconsistent background information. Third, even if we lived in a world with only such weird adversaries -- whether an adversary 'infers' about an individual would not always depend on whether that individual was actually ever included in data processing. For these reasons, the 05/2014 definition of inference is incompatible with the principles of Process Protection, Inclusion-Based Protection, and Format Neutrality.

The problems with this definition could not have been anticipated in the 1970s-1990s when the concepts underlying the 05/2014 definition were developed and refined. At the time, the field of formal privacy analysis was not developed sufficiently for the theoretical implications to be clear, and data releases were small and separated enough for the practical problems to be infrequent, and difficult to detect. However, given modern theoretical tools the problems are now clear, and in a world of ubiquitous data collection and frequent releases, the consequences of relying on a flawed definition are dire.

Defining inference so that it provides useful results, and meaningfully protects the people whose data is processed requires a more rigorously defined and targeted definition of inference. A regulation that regulates inference in the context of anonymization should clearly distinguish individual level inferences from population level inferences. This is needed both for making sure that useful statistics can be computed over the data (population level) and for providing regulatory protection for individual information. The regulation can use the inclusion principle to distinguish between the two - the principle identifies which information should be considered related to an individual, and this is the information which the regulation should protect.

The most widespread privacy-related measure that satisfies both process protection and inclusion is *counterfactual posterior-to-posterior risk.*[29]  All standard variants of DP are based on this metric, and satisfy the inclusion principle -- although the principle does not require the use of DP.[30]

---

[27] Shmueli, Erez, and Tamir Tassa. "Privacy by diversity in sequential releases of databases." Information Sciences 298 (2015): 344-372.
[28] Jarmin, Ron S., et al. "An in-depth examination of requirements for disclosure risk assessment." Proceedings of the National Academy of Sciences 120.43 (2023): e2220558120.
[29] See for a discussion Jarmin, Ron S., et al. "An in-depth examination of requirements for disclosure risk assessment." Proceedings of the National Academy of Sciences 120.43 (2023): e2220558120.

## 6. Singling Out: A Principled Analysis

In this section and the next we examine whether competing formal definitions for the GDPR privacy concepts of 'singling out' and 'linkability' align with the minimal formal requirements described above. We treat these two more briefly than inference because we have laid out foundations in the former section, and because we provide a detailed analysis of singling out in previous work (as noted below)

Opinion 05/2014 defines singling out as 'correspond[ing] to the possibility to isolate some or all records which identify an individual in the dataset'.[31] In prior work we argue that this definition equates singling out with (row) *isolation*.[32] Informally, a person is *isolated* in a dataset if they are described in a way that distinguishes them from all others in the dataset.

Prior work shows that some level of isolation is inevitable. An adversary can sometimes isolate a row in a dataset even without seeing any data release. Therefore, a regulation that requires isolation to be very unlikely or impossible violates **ALTER** (and **NJER**), as even the **Zero mechanism** cannot offer that guarantee. Since no mechanisms are permitted, such a regulation satisfies our remaining principles trivially, but not meaningfully.

### 6.1 K-anonymity

Opinion 05/2014 asserts that k-anonymity guarantees protection against singling out. This is inconsistent with the above interpretation of singling out as mere row isolation (as k-anonymity offers even less protection against isolation than the Zero mechanism[33]). Thus, one might consider an alternative interpretation of singling out, where k-anonymity suffices to prevent singling out. A regulation allowing k-anonymous mechanisms would satisfy ALTER (the Zero mechanism is vacuously k-anonymous) and NJER (as many k-anonymous algorithms have been proposed and deployed). The **identity mechanism** is not k-anonymous, satisfying **non-triviality.**

K-anonymity doesn't limit the informational relationship between input and output. Further, even as practiced, k-anonymity enables individual-level disclosure, under non-contrived assumptions about data distribution. The dominant class of k-anonymous algorithms work by efficiently suppressing attributes in the input microdata records (or by generalizing those attributes according to some data hierarchy). Typically, suppressing fewer attributes is desirable. Surprisingly, for some data distributions, algorithms that suppress as few attributes as possible universally allow a fraction of the microdata records to be completely recovered by a downstream analyst.[34] The same is true for stricter variants of k-anonymity, including l-diversity and t-closeness.

---

[30] For example, inclusion can potentially be satisfied through alternative definitions using the Pufferish framework. Kifer, Daniel, and Ashwin Machanavajjhala. "Pufferfish: A framework for mathematical privacy definitions." *ACM Transactions on Database Systems (TODS)* 39.1 (2014): 1-36.
[31] Aloni Cohen and Kobbi Nissim, 'Towards formalizing the GDPR's notion of singling out' (2020) 117(15) Proc. Natl. Acad. Sci., 8344-8352; Micah Altman, Aloni Cohen, Kobbi Nissim and Alexandra Wood, 'What a hybrid.
[32] Altman et al. (n 30), 15.
[33] Cohen, Aloni. "Attacks on Deidentification's Defenses." 31st USENIX Security Symposium (USENIX Security 22). 2022.

This has implications for **post-processing**.[35] It seems reasonable to hypothesize that a data protection regulation would restrict the publication of individual records in the clear. If so, then such a regulation must similarly restrict the use of k-anonymous algorithms, even those working by generalization and suppression. Not doing so would violate one of the **post-processing** or **non-contrived assumption** principles. The former because the post-processing of an unrestricted mechanism's output would be restricted, or the latter by tailoring assumptions on the data distribution to prevent the recovery of individual data records as described above.[36]

K-anonymity is defined in terms of the structure of a data release. As revealing the mechanisms does not affect a data release's structure, the **transparent complement** of permitted mechanisms are permitted. K-anonymity, however, violates **parallel composition**. Existing works show that k-anonymous mechanisms fail to compose in reasonable settings---settings permitted unless the non-contrived assumption principle is violated.[37]

A regulation that treated k-anonymity as sufficient to anonymize outputs in effect treats k-anonymized outputs as unrelated to any individual. It should also (see 3.3.1) treat non-k-nonymized outputs as related to an individual, if that individual is in D, and at least 1 but fewer than k rows in the output share the same quasi-identifiers of that individual. This regulation violates **individual relatedness**, because a k-anonymous output can nevertheless be substantially affected by the presence of a single individual. For example, in some circumstances individuals' data can be recovered from the output as described above.[38] It also violates individual unrelatedness, as an input record can have essentially no effect on the output while still sharing quasi-identifiers with one output row.[39]

---

[34] Aloni Cohen, "Attacks on Deidentification's Defenses" (2022) USENIX Security Symposium

[35] Less interestingly, k-anonymity violates post-processing in a "syntactic" sense. By redacting some number of records from a k-anonymized data release, a post-processor could violate the requirement that each collection of quasi-identifiers appears at least k times in the release. But a regulation might reasonably attempt to circumvent this issue by permitting k-anonymous outputs or redactions thereof.

[36] It is not yet known which distributions enable or prevent the downcoding attacks described in (Cohen 2022). So not only would the assumption needed to prevent these attacks be contrived, we know of no other way to even specify that assumption.

[37] Stokes, K., Torra, V.: Multiple releases of k-anonymous data sets and k-anonymous relational databases. Int. J. Uncertain. Fuzziness Knowl.-BasedSyst. 20(06), 839–853 (2012). https://www.worldscientific.com/doi/abs/10.1142/ S0218488512400260; Aloni Cohen, 'Attacks on Deidentification's Defenses' (2022) Proc of 31st USENIX Security Symposium; Srivatsava Ranjit Ganta, Shiva Kasiviswanathan, and Adam Smith, 'Composition Attacks and Auxiliary Information in Data Privacy' (2008) Proc of 14th ACM SIGKDD 265–273.

[38] Aloni Cohen, "Attacks on Deidentification's Defenses" (2022) USENIX Security Symposium

Such a regulation would not violate **format neutrality**.[40] A regulation *only* allowing k-anonymous data releases (at least, without data subjects' consent) would violate format neutrality. It would also violate **excluded individuals**, by prohibiting non-k-anonymous mechanisms whose outputs are unrelated to any individual (e.g., DP synthetic data).

## 6.2 Predicate singling-out

In prior work, we present an alternative called *predicate singling-out (PSO)*. Very roughly, a record is predicate singled-out in a dataset if the record can be described in a way that distinguishes it from all others in the dataset, and the description is so specific as to be unattributable to chance. A privacy mechanism prevents predicate singling-out attacks if the chance that an attacker manages to isolate a record in the dataset using an exceedingly rare predicate is small. Predicate singling-out captures the spirit of isolation while ruling out chance, making it a more appropriate concept for reasoning about privacy.

Clearly, the zero-mechanism prevents PSO attacks, and the identity mechanism does not. A hypothetical regulation permitting only mechanisms that prevent PSO attacks would thus satisfy **non-triviality** and **ALTER.** In prior work we present mechanisms that satisfy PSO, satisfying **NJER.** The PSO definition is constructed to satisfy **transparent complement** and we prove it satisfied **postprocessing** -- thus by implication it satisfies **format neutrality.** However, we show it does not satisfy **parallel composition.**

Similarly to k-anonymity and isolation, PSO is defined on rows, not individuals in the population, and thus does not inherently define individual relatedness. Following the approach in 3.3.1, we now consider a regulation that recognizes PSO-secure output as anonymised, that treating PSO-secure output as unrelated to any individual, and that treats non-anonymized output as related to an individual i iff individual *i* is measured in D, and the row corresponding to *i* can be singled out in O using an appropriate predicate.

This regulation satisfies excluded individual,[41], and individual unrelatedness. However, the regulation does not satisfy individual relatedness. A mechanism can output a small number of attributes from every input record without violating PSO security---enough so that every row in the output is distinct---so long as the attributes produced are not detailed enough to be very unlikely to occur by chance. The output depends significantly on each input *i*, but would not count as related to any of them by the above standard.[42]

---

[39] For example, consider the mechanism which: (1) discards the first record, then (2) generalizes the quasi-identifiers of the remaining records enough to cover all possible quasi-identifiers without overlapping. The first record has essentially no effect on the output, but shares quasi-identifiers with exactly one output row.

[40] K-anonymity being defined by the structure of the output, the conclusion should perhaps be that our approach to format neutrality could be strengthened.

[41] To be PSO secure (as we defined it), a mechanism only has to guarantee something for data sampled iid from a distribution. The hypothesis of UMD ("never a causal relationship") is stronger in that it holds for all data (and all auxiliary information).

## 7. Linkability: A Principled Analysis

Opinion 05/2014 defines linkability broadly as 'the ability to link, at least, two records concerning the same data subject or a group of data subjects (either in the same database or in two different databases)'.[43] No formal technical definition is provided, and the use of the term 'link' to define linkability has a degree of circularity. However, the term 'record linkage' has a long and broad usage in official statistics and the traditional statistical disclosure control literature deriving from the foundational work of Fellegi & Sunter (1969).[44]

**Double Matching.** Under the common definition which we refer to as "double match", a linkage occurs when: (a) two datasets, $D_1$ and $D_2$, comprise records measuring ($m_1$,$m_2$) of individuals (or other units of observation) drawn from the same underlying population P; (b) $D_1$ and $D_2$ contain some different measures ($m_1$ is not equal to or logically implied by, $m_2$); (c) and the attacker correctly identifies subsets of records $S_1$ and $S_2$ from $D_1$ and $D_2$, respectively, such that $S_1$ and $S_2$ represent identical subsets of units in P. That is, unit u is in $S_1$ if and only if u is in $S_2$. An individual record linkage occurs when the size of both subsets $S_1$ and $S_2$ is exactly 1. As correct linkages can be created through pure random mappings,[45] a linkage attack is said to succeed when the attacker can identify linkages at a rate significantly exceeding chance. We restrict our attention to **high-confidence** (deterministic) linkage, not probabilistic linkage: purported individual / group matchings must always be correct (or very nearly so).

**Double Jeopardy.** Although less uniformly defined and commonly there is a broader concept of linkage, with a history of at least thirty years, based on the risks of appearing in multiple datasets.[46] From this broader perspective, we can understand the harms commonly associated with data linkage arise when an adversary is able to learn substantially more about a specific individual because they were included in two different databases.

We describe double jeopardy formally, as follows:

**Double jeopardy linkability**

---

[42] We conjecture that PSO imposes some restriction on how much the output may disclose about individual inputs, but such a bound is not currently known.
[43] Opinion 05/2014 (n 2), 11-12.
[44] Ivan P. Fellegi and Alan B. Sunter, 'A theory for record linkage' (1969) 64(328) Journal of the American Statistical Association 1183-1210. See Duncan and Lambert, 1989, The Risk of Disclosure for Microdata, JBES.
[45] Ibid.
[46] See for example the discussion in section 5 of Duncan and Lambert, 1989, The Risk of Disclosure for Microdata, which models a success linkage attack as being measured by the amount that they learn (as a loss function) about an individual person in one database from a correct match as compared to a match to a random record.

> Let A, B, and C be disjoint sets of units in a population. Let dataset $D_1$=A+C and $D_2$=B+C consist of measures of the units in A+C and B+C, respectively. C are those units in common.
>
> We denote a mechanism's outputs on these datasets as $y_1=M(D_1)$, $y_2=M(D_2)$
>
> Also let there be a set of datasets with the common measured units removed:
>
> $\quad D_1' = D_1 - C = A, \; D_2' = D_2 - C = B$
>
> And denote the corresponding outputs as:
>
> $\quad y'_1 = M(D'_1), y'_2 = M(D'_2)$
>
> A linkage attack is successful under double jeopardy if for some function f:
>
> $\quad Pr(\, f(C) \mid y_1, y_2\,) \neq Pr(\, f(C) \mid y_1', y_2'\,)$ AND
>
> $\quad Pr(\, f(C) \mid y'_1, y'_2\,) \approx Pr(\, f(C) \mid y_1', y_2\,) \approx Pr(\, f(C) \mid y_1, y_2'\,)$

We now consider a regulation that treats processing by a mechanism that prevents linkability as sufficient for anonymization; treats anonymized output as unrelated to any individual (following 3.3.1); and treats an output $O_1=M(D_1)$ as related to individual *u*, if *u* is in $D_1$ and if an attacker can use $O_1$ to link $D_1$ to an external dataset $D_2$ with subsets $S_1$ and $S_2$ containing u.

The zero-mechanism guarantees that the output prevents both double-matching and double-jeopardy, and the identity mechanism guarantees protection from neither -- thus the regulation satisfies **ALTER** and **non-triviality** under either definition of linkability. **NJER** can be satisfied by either definition, for different reasons. For double jeopardy, **DP** mechanisms guarantee parallel composition across any number of releases, and are thus admissible. For double-matching, both DP and swapping and permutation mechanisms (such as the one illustrated in Example 1 below) can prevent individual record linkage.

No formal proofs establish whether mechanisms satisfying linkability of either type necessarily satisfy **transparent complement**. However, mechanisms such as those above exist that can remain safe under complement -- so a modified regulation explicitly requiring that any M used protect against linkability, *even if the adversary possesses knowledge of the mechanism* (non-vacuously) satisfies **transparent complement**. Similarly, since the mechanisms above achieve protection without relying on strong assumptions, a regulation requiring the use of mechanism that prevents linkability, even in the presence of adversarial knowledge would also satisfy **non-contrived assumptions**.

A regulation using the double-jeopardy definition of linkability also satisfies all remaining properties: **Post-Processing, format neutrality, parallel composition, individual relatedness, individual unrelatedness,** and **excluded individual** properties**.**

Surprisingly, however, a regulation using the double matching definition of record linkage violates Proces Protection. A hint as to how this happens is the implicit but common assumption that the output structure itself can protect against some forms of linkability. For example, Opinion 05/2014 states that k-anonymity reduces the risk of individual linkability to 1/k.[47] The intuition underlying this conclusion is that, although groups could be linked, the output grouping restricts further linkages; however, this intuition only makes sense if one focuses on the output alone and not the mechanism itself (nor the informational relationship between inputs and outputs that the mechanism provides). Example 1 below shows how defining linkability through permutation-matching violates the process principle.

> **Example 1.**
>
> *Consider a dataset containing information of all Members of the European Parliament. Each record in X contains name, nationality, and COVID status. We consider two algorithms for de-identifying the dataset.*
>
> *Algorithm A sorts the records by name, and replaces the names with pseudonyms — the numbers 1, 2, 3, etc. in order.*
>
> *Algorithm B has additional steps. First it sorts by name. Second, for every combination of nationality and COVID status, Algorithm B randomly shuffles all records that share that nationality and status. Third, Algorithm B replaces the names on all records with the numbers 1, 2, 3, etc. — just like Algorithm A, but after the shuffle step. Importantly, Algorithm B only shuffles records that share nationality and COVID status. Hence, it does not change, say, the 80th record's nationality or status, only the name associated with the 80th record before it is pseudonymized.*
>
> *Algorithm A would seem to enable individual linkage attacks, whereas Algorithm B would not. For instance, the one could immediately infer that record #1 corresponds to MEP Adamowicz of Poland.[48] In fact, one could perfectly re-assign names to every record.*
>
> *On the other hand, Algorithm B would seem to prevent individual linkage attacks.[49] There may be, say, 10 records that share the nationality and COVID status with record #1. It is impossible to know which of these records corresponds to MEP Adamowicz. While it is possible to infer MEP Adamowicz's COVID status by checking record #1, it would seem that perfect individual linkage is no longer possible.*
>
> *The above analysis of individual linkage under Algorithms A and B is very intuitive, and in line with common conceptions of record linkage.*

---

[47] Opinion 05/2014 (n 2), 16-17.
[48] https://en.wikipedia.org/wiki/List_of_members_of_the_European_Parliament_(2019%E2%80%932024)
[49] Footnote: Example 1 can be extended to group linkage for groups of size any size n. Algorithms A and B sort their input dataset, then apply k-anonymity with k > 2n. Algorithm B additionally randomly permutes the rows in each of the resulting "equivalence classes."

> *However, this intuitive analysis violates the Process Principle. This is because Algorithms A and B have identical input-output behavior. On every possible input, the output of Algorithm A and Algorithm B are exactly equal.*

Double-matching linkability does not satisfy parallel composition: For example, an algorithm that provides 2-anonymity prevents linking individuals with high-confidence in many circumstances. By construction any linkable group will have at least two members. However, as noted in section 3.3.1, k-anonymity is not composable. Another 2-anonymous release from the same dataset could enable individual-level double matching.

Double-matching linkability does satisfy the **Excluded Individual** principle. By the definition of double-matching, linkage amounts to specifying subsets of two datasets that correspond to the same individuals in the population. This requires any linked individuals to be contained in both datasets. Double-matching linkability also satisfies **Individual Unrelatedness**, for an appropriate choice of parameters**.** If the effect of individual *i* on the output of mechanism M is slight, then the risk of double-matching can only be slightly greater with *i*'s record than without it. But double-matching is impossible when *i*'s record is excluded. Hence the risk with *i*'s record included is also "slight." As double-matching requires the risk to "significantly exceed chance," double-matching is ruled out when an individual's effect on the mechanism is sufficiently "slight." However, double-matching linkability *does not satisfy* **Individual Relatedness** because a mechanism may prevent linkability but still be substantially affected by an individual inclusion in the data. For instance, a mechanism reporting exact means will be substantially affected by outliers, as can the equivalence classes of a k-anonymous mechanism.